\documentclass{aa}

\usepackage{graphicx}
\usepackage{txfonts}
\usepackage[T1]{fontenc}
\usepackage{bm}
\usepackage{physics}
\usepackage{amsmath}	
\usepackage{booktabs}
\usepackage{multirow}
\usepackage{color}
\usepackage{supertabular}
\usepackage{subcaption}
\usepackage{hyperref}

\usepackage{ulem}
\newcommand{\HFR}{\texttt{HFR}}
\def\Msun{$M_{\odot}$}
\def\ga{\,\,\raise0.14em\hbox{$>$}\kern-0.76em\lower0.28em\hbox
{$\sim$}\,\,}
\def\la{\,\,\raise0.14em\hbox{$<$}\kern-0.76em\lower0.28em\hbox
{$\sim$}\,\,}

\begin{document}

\title{Kilonova ejecta opacity inferred from new large-scale HFR atomic calculations in all elements between Ca ($Z=20$) and Lr ($Z=103$)\thanks{The complete atomic and opacity data sets used in this paper are available on the Zenodo database at the address \url{https://zenodo.org/records/14017953} (DOI 10.5281/zenodo.14017952).}}


   \author{J. Deprince
          \inst{1,2}
          \and
          G. Wagle\inst{2}
          \and
          S. Ben Nasr\inst{1}
          \and 
          H. Carvajal Gallego\inst{1}
          \and 
          M. Godefroid\inst{3}
          \and \\
          S. Goriely\inst{2} 
          \and 
          O. Just\inst{4,5}
          \and
          P.~Palmeri\inst{1}
          \and
          P.~Quinet\inst{1,6}
          \and
          S. Van Eck\inst{2}
          }

   \institute{Atomic Physics and Astrophysics, Universit\'e de Mons (UMONS), Mons (Belgium)
         \and
              Astronomy and Astrophysics Institute, Universit\'e libre de Bruxelles, Brussels (Belgium)
         \and
             Spectroscopy, Quantum Chemistry and Atmospheric Remote Sensing, Universit\'e libre de Bruxelles, Brussels (Belgium) 
        \and 
            GSI Helmholtzzentrum für Schwerionenforschung, Darmstadt (Germany)
        \and
            Astrophysical Big Bang Laboratory, RIKEN Cluster for Pioneering Research, Wako (Japan)
        \and 
            IPNAS, ULiège, Liège (Belgium)
            }

   \date{}

 
  \abstract
   {The production of elements heavier than iron in the Universe still remains an unsolved mystery. About half of them are thought to be produced by the astrophysical r-process (rapid neutron-capture process), for which one of the most promising production sites are neutron star mergers (NSMs). In August 2017, gravitational waves generated by a NSM were detected for the first time by the LIGO detectors (event GW170817), and the observation of its electromagnetic counterpart, the kilonova (KN) AT2017gfo, suggested the presence of heavy elements in the KN ejecta. The luminosity and spectra of such KN emission depend significantly on the ejecta opacity. Atomic data and opacities for heavy elements are thus sorely needed to model and interpret KN light curves and spectra.}
   {The present work focusses on large-scale atomic data and opacity computations for all heavy elements with $Z \geq 20$, with a special effort on lanthanides and actinides, for a grid of typical KN ejecta conditions (temperature, density and time post-merger) between one day and one week after the merger (corresponding to the LTE photosphere phase of the KN ejecta).}
   {In order to do so, we used the pseudo-relativistic Hartree-Fock (HFR) method as implemented in Cowan’s codes, in which the choice of the interaction configuration model is of crucial importance.}
   {In this paper, HFR atomic data and opacities for all elements between Ca ($Z=20$) and Lr ($Z=103$) are presented, with a special focus on lanthanides and actinides. Besides, we also discuss the contribution of every single element with $Z \geq 20$ to the total KN ejecta opacity for a given neutron star merger model, depending on their Planck mean opacities and elemental abundances. An important result is that lanthanides are found to not be the dominant sources of opacity, at least on average. The impact on KN light curves of considering such atomic-physics based opacity data instead of typical crude approximation formulae is also evaluated. In addition, the importance of taking the ejecta composition into account directly in the expansion opacity determination (instead of estimating single-element opacities) is highlighted. A database containing all the relevant atomic data and opacity tables has also been created and published online along with this work.}
   {}


   \maketitle
%

\section{Introduction}

The stable neutron-rich nuclides heavier than iron observed in stars of various metallicities and in the solar system are produced by the rapid neutron-capture process (or r-process) of stellar nucleosynthesis.
The r-process is responsible for creating about half of the atomic nuclei heavier than iron. This process occurs in explosive environments with extremely high neutron densities and temperatures \citep{Arnould07,Arnould20,Cowan21}. The exact astrophysical sites of the r-process have been extensively searched. For a long time, core-collapse supernovae of massive stars were considered the primary r-process sites, but, so far, no successful r-process has been found in the most sophisticated models \citep[e.g.][and references therein]{Janka17,Wanajo18a,Wang23}. A subclass of core-collapse supernovae, known as collapsars and corresponding to rapidly rotating and highly magnetized massive stars and associated with the origin of observed long $\gamma$-ray bursts remain promising r-process sites \citep{Winteler12,Siegel19b,Miller19d,Just22a}. Alternatively, neutron star (NS) mergers have been also envisioned as a potential site for the r-process \citep{Lattimer74,Eichler89}. Hydrodynamical simulations of NS mergers have confirmed that a considerable amount of matter could be ejected from these systems \citep[e.g.][]{Ruffert96,Rosswog99}. Associated nucleosynthesis calculations have shown that NS mergers are viable r-process sites \citep[see e.g.][]{Goriely11b, Korobkin12, Wanajo14,Just15,Radice18,Kullmann23,Just2023}. 
The resulting abundance distributions are found to reproduce remarkably well the solar system distribution, as well as various elemental distributions observed in low-metallicity stars. In addition, the ejected mass of r-process material, combined with the predicted astrophysical event rate (around 10\,My$^{-1}$ in the Milky Way) can account for the majority of r-process material in our Galaxy.  Another piece of evidence that NS mergers are major contributors to the Galactic r-process-enrichment comes from the remarkable 2017 gravitational-wave event GW170817 \citep{Abbott17} and its electromagnetic counterpart AT 2017gfo \citep{Abbott17d}. The decay of freshly produced radioactive r-process elements is expected to power such a kilonova (KN) which is potentially observable in the optical and (near-)infrared electromagnetic spectrum. Fits to the observed KN light curve provided a strong indication that r-process elements have been synthesised in this event \citep[e.g.][]{Kasen17}. \citet{Watson19,Gillanders22} also reported the spectroscopic identification of Sr ($Z=38$), providing the first direct evidence that NS mergers eject material enriched in heavy elements.  

The light curve of a KN delivers unique information about the ejecta mass and velocity, and its composition. However, the modeling of these merger events and the resulting KN explosion still face considerable challenges, including in particular the description of radiative processes that determine the opacity and emission observed. This requires the most complete knowledge possible of atomic data such
as energy levels, transition wavelengths, and oscillator strengths characterizing the species contributing to the opacity. A major effort has been made in this direction in recent years with various studies aimed at determining the relevant atomic parameters and deducing the corresponding opacity for many heavy elements. Among these studies, we will mention the works published by \citet{Tanaka2020} for elements ranging from Fe to Ra in ionization stages from I to IV, by \citet{Fontes2020,Fontes2023} for all lanthanide and actinide atoms in charge stages between I and IV, by \cite{Banerjee2022} for three selected lanthanides (i.e., Nd, Sm and Eu) from V to XI, and by \citet{Banerjee2024} for all elements from La to Ra in the charge stages ranging from I to XI, to which must be added investigations more focused on specific elements such as the works published by \citet{Gaigalas2022} for Pr IV, \citet{Gaigalas2019} for Nd II–IV, \citet{Gaigalas2020} and \citet{Deprince2024} for Er III, \citet{Radziute2020} for Pr II–Gd II, \citet{Radziute2021} for Tb II–Yb II, \citet{Carvajal2021} for Ce II–IV, \citet{Rynkun2022} for Ce IV, \citet{Carvajal2022a} for Ce V–X, \citet{Carvajal2022b} for La V–X, \citet{Carvajal2023a} for Pr V–X, Nd V–X, Pm V–X, \citet{Carvajal2023b} for Sm V–X, \cite{Maison2022} for Lu V, \cite{BenNasr2023} for Nb I-IV, Ag I-IV, \citet{BenNasr2024} for Hf I-IV, Os I-IV, Au I-IV, \citet{Deprince2023} for U II-III and \citet{Flors2023} for Nd II-III and U II-III. Let us also add to all these investigations, the recent paper published by \citet{Peng2024} concerning KN light curve interpolation with neural networks.

In the present work, we report new opacity calculations for all elements with $20 \leq Z \leq 103$ (\textit{i.e.} from Ca to Lr), using a consistent set of atomic parameters obtained with the pseudo-relativistic Hartree-Fock (\HFR) method for all atomic species of interest in the first four charge states (I -- IV). A comparison with the most complete previous works is also discussed, as well as the impact on kilonova light curve models.

In Sect.~\ref{sec:hfr}, the \HFR\ method is described, as well as the atomic models used for each ion considered in this work. A discussion about the accuracy of our atomic data is also carried out. The expansion opacities of all elements from Ca to Lr are illustrated for typical KN ejecta conditions in Sect.~\ref{sec:exp-opa}, as well as a comparison with a similar previous work. The same type of comparison is performed in Sect.~\ref{sec:LB} concerning the Planck mean line-binned opacities. In Sect.~\ref{sec:database}, we introduce and detail the new HFR atomic database and opacity tables produced in this work and available online. Finally, in Sect.~\ref{sec:astro}, the impact of the newly calculated \HFR\ opacities is illustrated in the case of a NS merger simulation. A special attention is paid to the elements contributing the most to the ejecta opacity and to their impact on the light curve in comparison with alternative approximations.


\section{HFR computations}
\label{sec:hfr}
The \texttt{pseudo-Relativistic Hartree-Fock} code (\HFR) was developed by \citet{Cowan1981}. In this method, a set of orbitals is obtained for each configuration by solving the Hartree-Fock (HF) equations that are derived from the application of the variational principle on the corresponding configuration average energy. In addition, several relativistic corrections are included in a perturbative way, namely the Blume-Watson spin-orbit (including the one-body Breit interaction operator), mass-velocity and one-body Darwin terms. The coupled HF equations are solved by using the self-consistent field approach.

Within the framework of the Slater-Condon method, the atomic wavefunctions (eigenfunctions of the Hamiltonian) are built as a superposition of basis wavefunctions in the $LSJ\pi$ representation, \textit{i.e.} 
\begin{equation}
\label{eq:CSF_exp_HFR}
    \Psi(\gamma J M_J \pi)= \sum_i^{N_{\texttt{CSF}}}c_i \; \phi(\gamma_i L_i S_iJ M_J \pi).
\end{equation}
The construction and diagonalization of the multiconfiguration Hamiltonian matrix are carried out within the framework of the Slater-Condon theory. Each matrix element is computed as a sum of products of Racah angular coefficients ($v^l_{ij}$), and radial Slater and spin-orbit integrals ($x_l$): 
\begin{equation}
    \mel{i}{H}{j} = \sum_l v^l_{ij} \; x_l.
\end{equation}
As recommended by \citet{Cowan1981}, scaling factors of 0.85 are applied to the Slater integrals in the present calculations. The choice of scaling factors between 0.8 and 0.95 virtually does not affect the computed expansion opacities as recently shown by \citet{Carvajal2023a} in the case of Nd IX. 

The radiative transition wavelengths and oscillator strengths for each transition allowed in the electric dipole approximation can then be computed from the eigenvalues and eigenstates obtained using such approach. Note that only the transitions characterized by $\log gf \geq -5$ are considered in this work, since it was shown by \citet{Carvajal2022a} that values under this cut-off do not modify the computed expansion opacity. \\

For all lanthanide and actinide ions considered in this work, namely from the neutral to the trebly-ionized species, the same methodology was adopted to build the multiconfiguration models, considering single and double electron substitutions from reference configurations. In more details, for all the lanthanides, the model for each ion was built by considering single electron excitations from the ground configuration towards all $n=5$ and $n=6$ orbitals, as well as configurations arising from double excitations from the ground one to selected $n=5$ and $n=6$ orbitals, namely 5d, 6s and 6p, with a few isolated restrictions for ions characterized by an atomic structure with a very large number of states (and thus by a very large Hamiltonian matrix to be diagonalized).
Similarly, for actinides, the same single and double excitations were considered to the corresponding $n=6$ and $n=7$ orbitals (\textit{i.e.} all the single electron substitutions from the ground configuration to $n=6$ and $n=7$ orbitals as well as double excitations towards 6d, 7s and 7p).
The lists of the configurations included in our models are given in the Appendix in Table \ref{table:config}. The justification of such model choices is based on the same strategy that we used for \ion{Nd}{ii}, \ion{Nd}{iii}, \ion{U}{ii} and \ion{U}{iii} in \citet{Flors2023}, in which we show that the resulting \HFR\ atomic data  lead to converged expansion opacities.  
The accuracy of the atomic data obtained for \ion{Nd}{ii}, \ion{Nd}{iii}, \ion{U}{ii} and \ion{U}{iii} have already been discussed in \citet{Flors2023}. In summary, in this paper, we compared our HFR atomic levels with the ones computed by another independent method, namely the Flexible Atomic Code (FAC) \citep{Gu2008}, and found an overall agreement between 10\% and 20\% for these four ions. When compared to experimental energy levels from the pioneer compilation of  \citet{Blaise1992} entitled "Selected Constants Energy Levels and Atomic Spectra of Actinides" (abbreviated as SCASA and available in an online database \citep{SCASA}), our \HFR\ energy levels agree within 18\% to 30\%, except for \ion{Nd}{iii} for which a larger discrepancy was obtained (49\%), doing in some cases worse and in other cases better than the FAC computations depending on the ion. Note that we also used the same strategy to build the \HFR\ multiconfiguration model of \ion{Er}{iii} in the same context in \citet{Deprince2024}, for which an overall agreement of 37\% was found when comparing our energy levels with experimental data from \citet{Wyart1997}. We showed in \citet{Deprince2023} and \citet{Deprince2024} that, even if a non-negligible discrepancy was obtained between our \HFR\ atomic levels and experimental data for \ion{U}{ii}, \ion{U}{iii} and \ion{Er}{iii}, a calibration of our data consisting in fitting the computed configuration average energies to the ones deduced from SCASA \citep{SCASA} and from \citet{Wyart1997} greatly improves the accuracy of our energy levels (\textit{e.g.}, the overall agreement was reduced from 37\% to 4\% in the case of \ion{Er}{iii}, see \cite{Deprince2024}) but has only a minor impact on the expansion opacities. In particular, the ground levels computed without calibration do not match the observed ones for those three ions (which is the case for 23 ions among all the 120 lanthanide and actinides ions considered in this work). However, as mentioned above, we observed that the expansion opacities were virtually not affected by the level inversion calibration and concluded that, in the framework of large-scale computations of atomic data for opacity determination, this calibration could be omitted --at least in a first step-- for ions with high spectral densities as lanthanides and actinides and for the conditions expected in KN ejecta \citep{Deprince2023,Deprince2024}. \\

For all the other elements from $Z= 20$ (\textit{i.e.} $20 \leq Z \leq 56$ and $72 \leq Z \leq 88$), namely those belonging to the fourth, fifth, sixth and seventh rows of the Periodic Table, we also considered all the neutrals and ions up to IV ionization stage. \\
Firstly, we focus on the elaboration of the configuration lists for the fourth row transition metal elements, namely with an open 3d subshell. For Co I and the corresponding isoelectronic fourth row ions (namely Ni~II, Cu~III and Zn~IV), which are all homologous to Ag~III, we used a configuration interaction (CI) model of the same type as the one used in our paper \citet{BenNasr2023}, i.e. with an [Ar] core instead of a [Kr] core and 3d up to 6f valence orbitals instead of 4d up to 8f valence orbitals. For the other neutral fourth row elements and their isoelectronic ions, the same list of configurations as for Co I were considered increasing or decreasing the number of electrons in the 3d subshell. For Sc~IV and the corresponding isoelectronic fourth row transition metal Ti V, single excitations of 3p electron from the 3p$^{6}$ configuration to valence subshells up to 6f were used to generate the list of configurations. As far as calcium ions are considered, for Ca~I, all single excitations of an electron 4s from the ground configuration 4s$^{2}$ to the valence subshells up to 6f were used to build the CI list which was further extended by adding the 3d$^{2}$ and 4p$^{2}$ configurations. Concerning the other ions, namely Ca~II, Ca~III and Ca~IV, which are respectively isoelectronic to Sc~III, Sc~IV and Sc~V, the same lists of interacting configurations were thus considered except for Ca~IV, for which the configuration 3s3p$^6$ was added in the list. The fourth row elements from $Z = 31$ to $Z = 36$ are characterized by the filling of the 4p subshell. For Ga~I, single excitations of the 4p electron from the ground configuration 4s$^{2}$4p to the valence orbitals up to 6f were considered to build the CI model in a first step. The list was then extended by adding the 4s4p$^{2}$, 4s4p5s, 4s4p4d and 4p$^{3}$ configurations. For all the other neutral species, the lists of interacting configurations were deduced by gradually filling the 4p subshell from the Ga~I CI model. Regarding the other ionization stages, the CI lists were deduced from the corresponding isoelectronic sequence of each ion (e.g., since Ge~II, As~III and Se~IV correspond to the isoelectronic sequence of Ga, we used the same multiconfiguration model for these ions).\\
Concerning the fifth-row elements ($Z = 37-54$), the lists of interacting configurations included in the various models were deduced from the corresponding fourth-row homologous ions ($Z = 20 - 36$), besides the neutral rubidium (since we considered elements from $Z=20$ for the fourth row) whose CI list was generated by considering single excitation of the 5s electron from the ground configuration 5s to valence subshells up to 7f. 
The same approach was also used to build the multiconfiguration models of the sixth-row elements ($Z = 72 - 86$) as well as for cesium ($Z=55$), barium ($Z=56$), francium ($Z=87$) and radium ($Z=88$), i.e. the configuration lists were built based on the homologous fifth-row element ones.\\
The configuration models used for all the elements considered in this work (\textit{i.e.} elements with $20 \leq Z \leq 103$) are presented in the Appendix in Table \ref{table:config}.

\section{Expansion Opacity}\label{sec:exp-opa}

\begin{figure}
    \centering
    \includegraphics[width=\hsize]{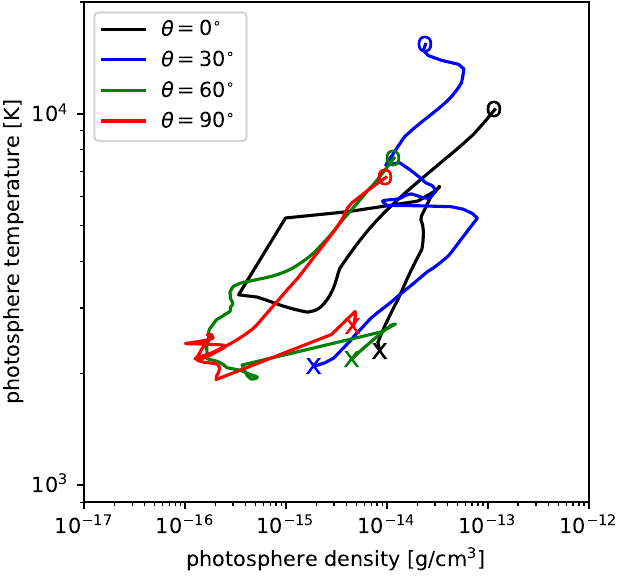}
    \caption{Time evolution of the density and temperature at the photosphere in KN ejecta of an exemplary NS merger simulation (model sym-n1-a6 of \citet{Just2023} with an ejecta mass of 0.073~\Msun) at four different polar angles $\theta$ relative to the rotation axis of the NS binary. Each line starts at $t=0.1$~d (indicated by circles) and ends when the radial optical depth drops below unity (indicated by crosses).}
    \label{fig:grid}
\end{figure}

We computed the bound-bound opacities of all the elements mentioned in Section \ref{sec:hfr} for typical conditions expected in the KN ejecta (roughly guided by the AT2017gfo case) from 1 to 7 days after the neutron star merger, which corresponds to the photosphere phase of the KN, in which local thermodynamical equilibrium (LTE) conditions are expected to be valid \citep{Pognan2022}. From roughly one week post merger, non-LTE (NLTE) effects are expected to be more important. In this study, we thus limit ourselves to the KN photosphere phase from 1 day after merger. The corresponding conditions of temperature and density within the KN ejecta can be estimated from ejecta distributions resulting in hydrodynamical simulations, as illustrated in Fig. \ref{fig:grid} showing the temperature and density conditions expected in the photosphere of the KN ejecta of a 1.375-1.375~\Msun\ NS merger \citep[model ``sym-n1-a6'' in][]{Just2023} from $t=0.1$~d until the ejecta becomes optically thin, and for which we adopted the heuristic prescription described in \cite{Just22b} for the opacity and photosphere. Based on these estimates, we compute the atomic opacities of all the elements between $Z=20$ and $Z=103$ for a grid of conditions corresponding to temperatures $1000$ K $\leq T \leq$ $10~000$ K and densities $10^{-17}$ cm$^{-3}$ $\leq \rho \leq 10^{-13}$ cm$^{-3}$. For such temperatures, only the low ionization stages of the elements are present in the KN ejecta, namely from the neutral to the trebly-ionized species \citep[see \textit{e.g.}][]{Tanaka2020,Flors2023}. We thus focus our work on these ions, already mentioned in Section \ref{sec:hfr}.
As in previous studies \citep[e.g.,][]{Flors2023, Tanaka2020}, the opacities given in this section (as well as in Section \ref{sec:LB}) serve as illustration and as means of comparison with data provided by other groups since they adopt a a composition with 100\% of a given element, which is unrealistic since the KN ejecta total opacity obviously depends on the elemental abundance pattern resulting from r-process nucleosynthesis. More realistic opacities based on ejecta compositions deduced from astrophysical simulations will be discussed in Section \ref{sec:astro}.

In this study, we mainly focus on the widely used expansion opacity \citep[see ][]{Eastman1993,Kasen2013,Tanaka2013} defined as
\begin{equation}
    \kappa_{\mathrm{exp}} (\lambda) = \frac{1}{ct\rho}\sum_l \frac{\lambda_l}{\Delta\lambda}\left(1-e^{-\tau_l}\right),
    \label{equ:expansion_opacity}
\end{equation}
where $t$ is the time after the merger, $\rho$ is the ejecta density, $\Delta\lambda$ is the wavelength bin width and $\lambda_l$ is the transition wavelength with Sobolev optical depth
\begin{equation}
    \tau_l = \frac{\pi e^2}{m_e c } f_l n_l \lambda_l t ,
    \label{equ:sobolev_optical_depth}
\end{equation}
in which $f_l$ is the oscillator strength of the line with transition wavelength $\lambda_l$ and the corresponding lower level number density $n_l$. The number densities of the lower levels can be computed under the LTE assumption by using the Saha ionization balance
\begin{equation}
    \frac{n_i}{n_{i-1}} = \frac{U_i(T)}{U_{i-1}(T)n_e}\frac{2}{\lambda_{\text{dB}}^3}e^{-E_{\mathrm{ion}}/k_B T}
\end{equation}
and Boltzmann population equations

\begin{equation}
    n_k = \frac{g_k}{U_i(T)}e^{-E_k/k_B T}n_i, 
    \label{equ:boltzmann_excitation}
\end{equation}
where the indices $i$ and $k$ indicate ionization stages and level numbers, respectively, $U_i(T)$ is the partition function, $n_e$ is the electron density, $\lambda_{\text{dB}}$ is the thermal de Broglie wavelength, $E_{\mathrm{ion}}$ is the ionization energy, $g_k$ is the multiplicity of the level $(2J_k+1)$ and $E_k$ is the level energy. 

In this work, the bin width is chosen to be 1\% of the wavelength, but it is worth mentioning that neither the absolute scale nor the general shape of the expansion opacity depends on the bin width, as mentioned e.g. by \citet{Flors2023}. In the literature, a bin width of $10\,$\AA\,is often assumed, but our choice results in smoother (and therefore easier to read) curves. The choice of using the Sobolev approximation \citep{Sobolev1960} to compute opacities in the expansion formalism is justified for NS merger ejecta \citep[see \textit{e.g.}][]{Flors2023} by the high expansion velocities ($\sim 0.1c$) of the ejecta.

The opacity thus depends on the physical conditions within the ejecta for a given time post merger, \textit{i.e.} temperature and density. To facilitate comparisons with other studies, we chose to illustrate our expansion opacities in this Section for typical conditions of the ejecta around 1 day after the merger, corresponding to a temperature of $T=5000$\,K (in accordance with the continuum of AT2017gfo inferred from the spectrum at 1.4\,days) and a density of $\rho=10^{-13}$\,g\,cm$^{-3}$ (characteristic for an ejecta mass of $\sim 10^{-2}$\,$M_\odot$ distributed uniformly within a sphere expanding at $0.1c$). 

For each ion considered in this work, we also computed the corresponding wavelength-independent Planck mean opacity, which is defined as 
\begin{equation}
    \kappa_{\mathrm{Planck}} = \frac{\int_0^\infty B_\lambda(T)\kappa_\mathrm{exp}(\lambda)d\lambda}{\int_0^\infty B_\lambda(T)d\lambda} .
    \label{eq:planck}
\end{equation}
Wavelength-independent (i.e. "gray") opacities are often used for light curve modeling. The use of Planck mean opacities is supported by the circumstance that the observed radiation in AT2017gfo closely resembled a blackbody spectrum \citep{Watson19,Gillanders22}.

\begin{figure}[!htb]
    \centering
    \includegraphics[width=\hsize]{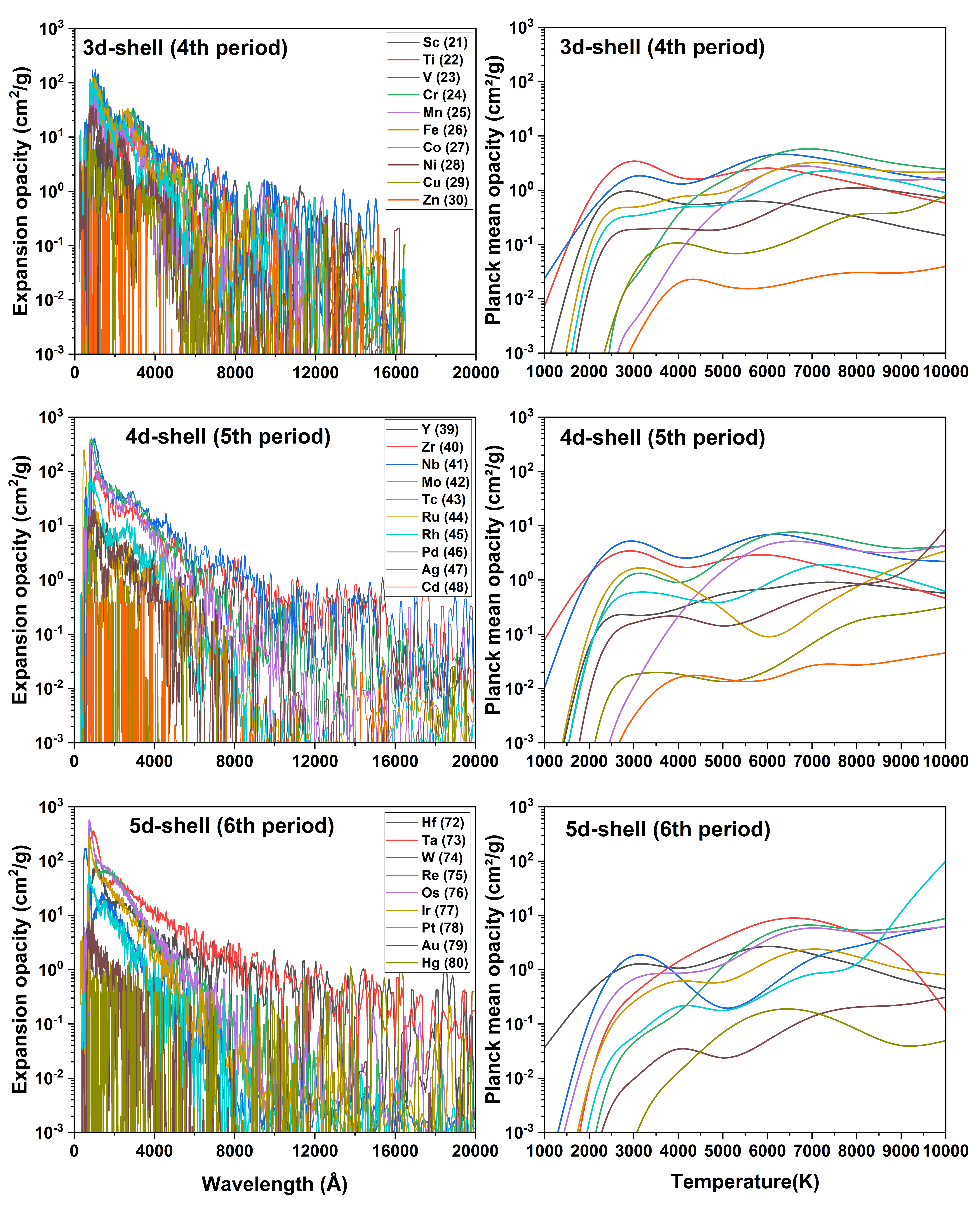}
    \caption{Expansion opacities of the $n$d group elements for $t=1$ day, $T=5000$K and $\rho=10^{-13}$ cm$^{-1}$ (left panels) and their corresponding Planck mean opacities (right panels).}
    \label{fig:nd}
\end{figure}

\begin{figure}[!htb]
    \centering
    \includegraphics[width=\hsize]{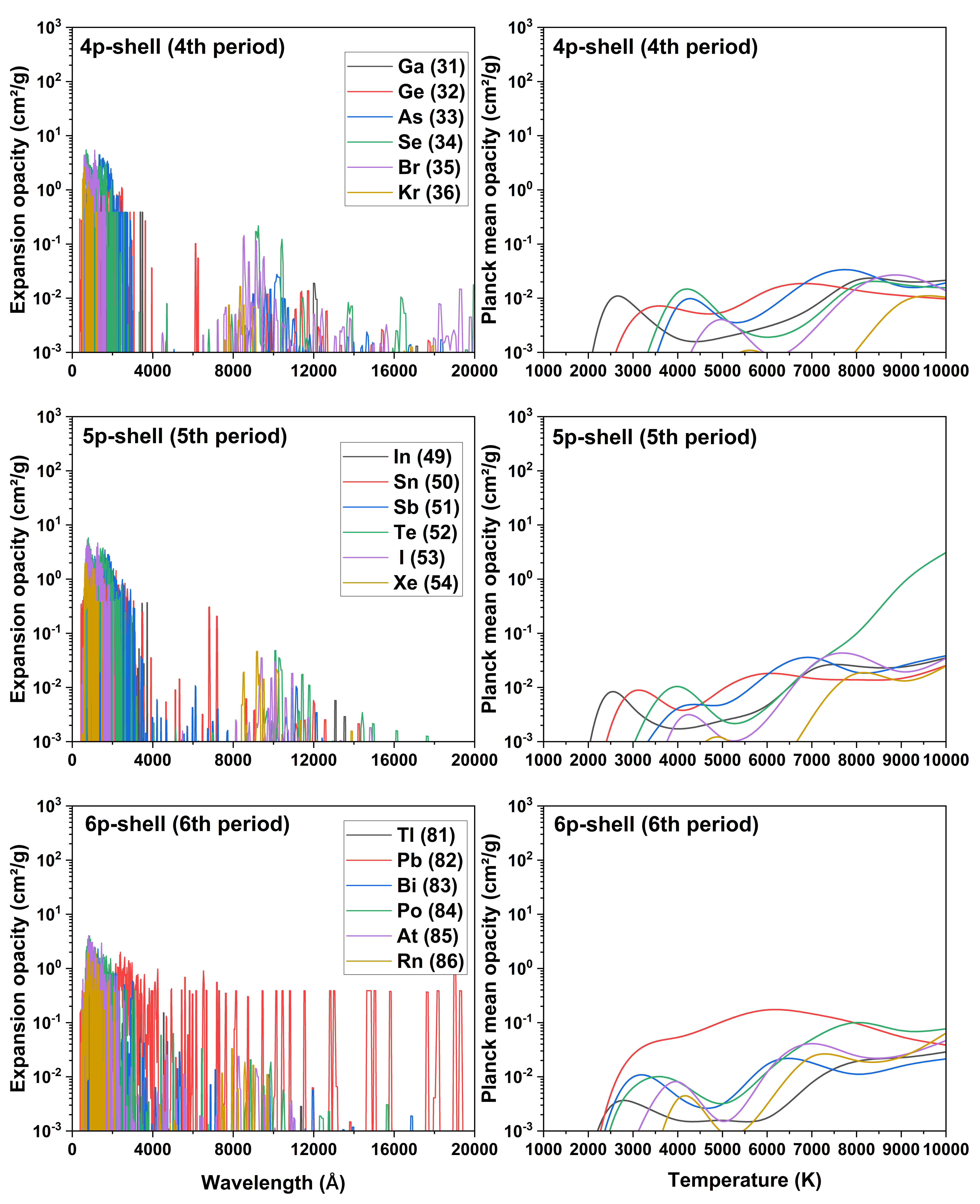}
    \caption{Expansion opacities of the $n$p group elements for $t=1$ day, $T=5000$K and $\rho=10^{-13}$ cm$^{-1}$ (left panels) and their corresponding Planck mean opacities (right panels).}
    \label{fig:np}
\end{figure}

\begin{figure}[!htb]
    \centering
    \includegraphics[width=\hsize]{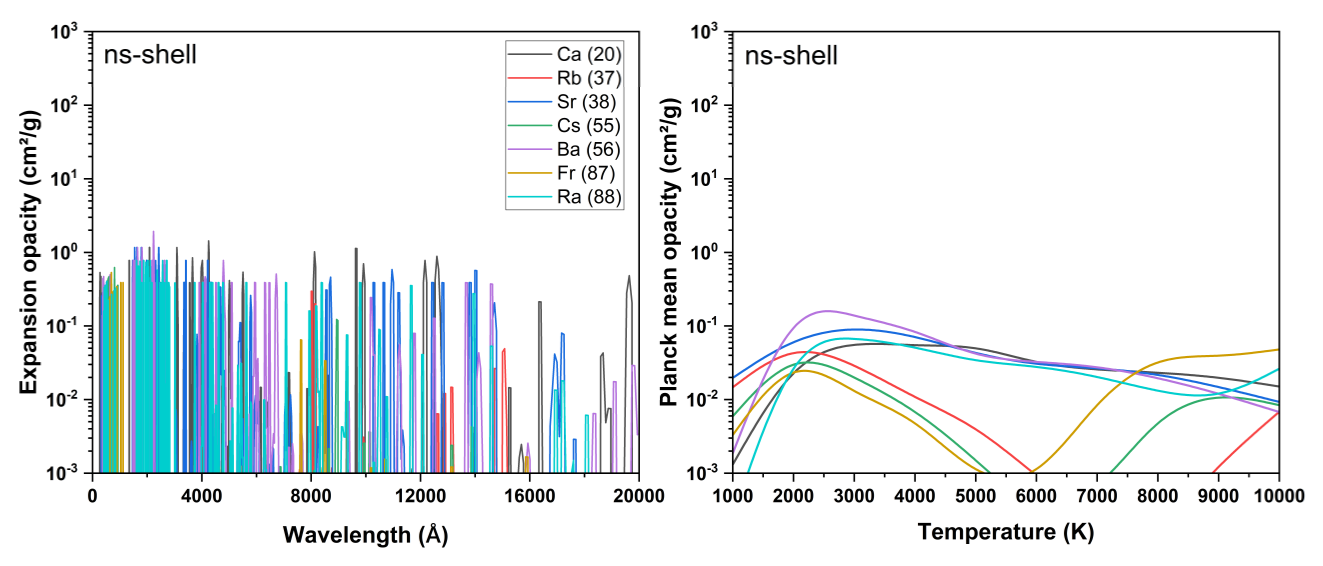}
    \caption{Expansion opacities of the $n$s group elements for $t=1$ day, $T=5000$K and $\rho=10^{-13}$ cm$^{-1}$ (left panels) and their corresponding Planck mean opacities (right panels).}
    \label{fig:ns}
\end{figure}

\begin{figure}[!htb]
    \centering
    \includegraphics[width=\hsize]{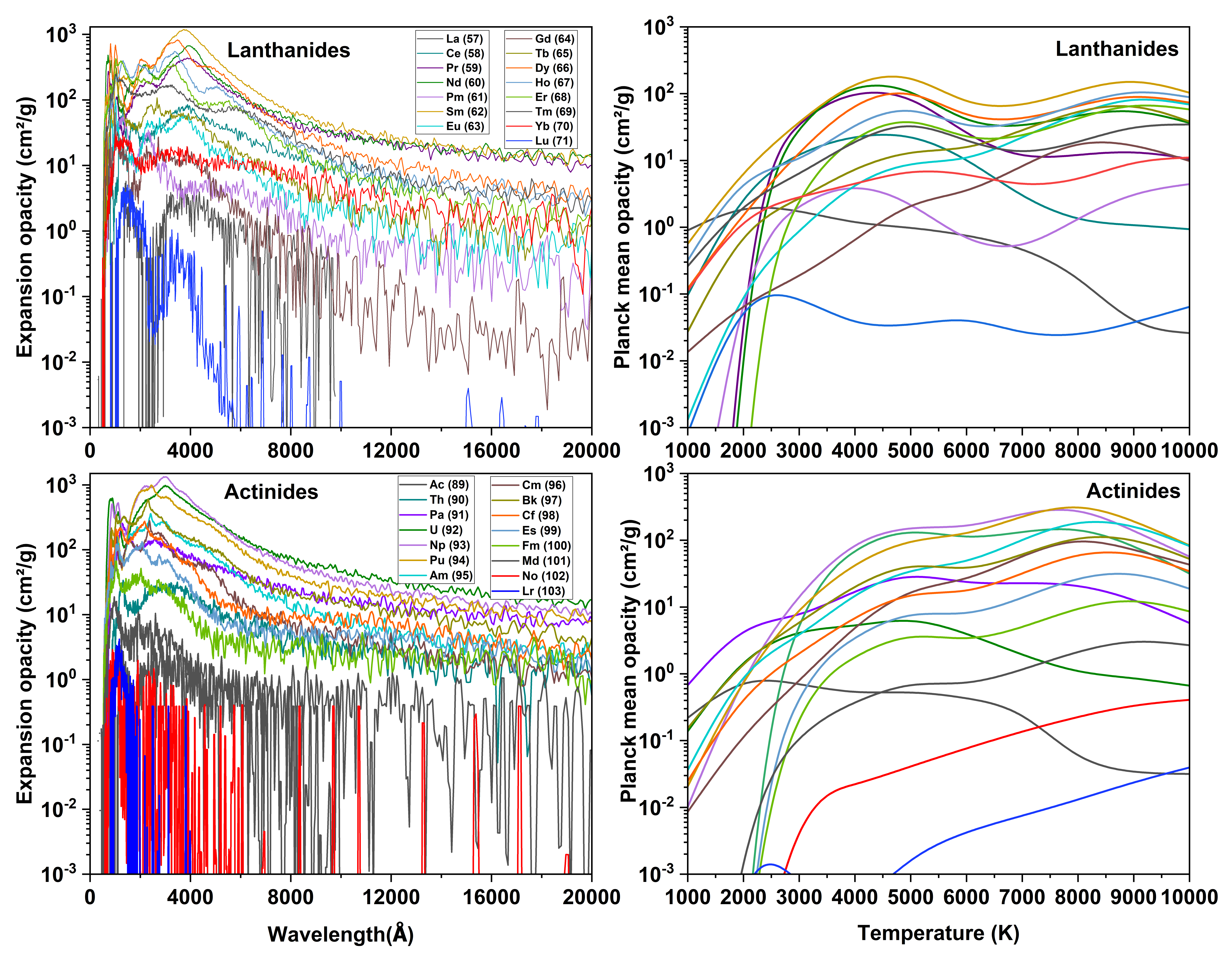}
    \caption{Lanthanide and actinides expansion opacities for $t=1$ day, $T=5000$K and $\rho=10^{-13}$ cm$^{-1}$ (left panels) and their corresponding Planck mean opacities (right panels).}
    \label{fig:lanthanides_actinides}
\end{figure}

The expansion opacities (left panels) computed for elements of the $n$d-shell group, $n$p-shell group, $n$s-shell group and for all lanthanides and actinides for the aforementioned conditions as well as their corresponding Planck mean opacities as a function of the temperature (right panels) are respectively shown in Figures~\ref{fig:nd}, \ref{fig:np}, \ref{fig:ns} and \ref{fig:lanthanides_actinides}. Once again, we want to stress that the discussion carried out in this section concerning the predominance of given elements with respect to its expansion opacity assumes a single-element plasma of each species, while a discussion about the contribution of each element to the total KN ejecta opacity based on the physical conditions and on realistic elemental abundances within the ejecta will be conducted in Section \ref{sec:astro}.

The d-shell element expansion opacities (see Figure~\ref{fig:nd}) exhibit a strong wavelength dependency, since they are higher at shorter wavelengths (smaller than 3000 \AA). Among the 3d-shell elements, V ($Z=23$) and Ti ($Z=22$) dominate the expansion opacity at short wavelengths (in the UV). The Planck mean opacity of these two elements gradually increases with temperature, reaching a peak of 30 cm$^2$/g and 50 cm$^2$/g, respectively, at $T=6000$ K. For $T=7000$ K, Cr ($Z=24$) has the highest opacity, whereas Zn ($Z=30$) has the lowest contribution to the opacity at any temperature.
In the 4d-shell series, the dominant elements are Nb ($Z=41$) and Mo ($Z=42$), reaching a maximum expansion opacity of 400 cm²/g in the UV, which rapidly decreases with increasing wavelengths. 
For intermediate temperatures ($T = 5000$ -- $8000$ K) Nb and Mo contribute the most to the expansion opacity, while Pd ($Z=46$) become predominant at higher temperature ($T=10000$ K). The species with the lowest expansion opacity is Cd ($Z=48$) for any temperature. As far as the 5d-shell group is concerned, W ($Z=74$) and Hf ($Z=72$) are the strongest contributors to the opacity at low temperatures, while Ta ($Z=73$), Os ($Z=76$) and Re ($Z=75$) become dominant at higher temperatures, as well as Pt ($Z=78$) above $T=9000$ K. Similarly to Zn and Cd from the 3d- and 4d-shell groups, Hg ($Z=80$) has the lowest expansion opacity within the 5d-shell group.

The elements from the p-shell group (see Figure~\ref{fig:np}) contribute less to the opacity in comparison to the d-shell elements. The highest opacities are observed in the UV. Within the 4p-shell group, all elements show a similar trend, no element clearly dominates. However, among the 5p-shell elements, Te ($Z=52$) contributes the most to the opacity at high temperatures. As far as the 6p-shell group is concerned, Pb ($Z=82$) dominates the opacity for almost any temperature.

The opacity of the s-shell group elements (see Figure~\ref{fig:ns}) are nearly insignificant compared to the p-shell and d-shell ones, in light of the limited number of transitions possible in such elements due to their atomic structure characterized by much less energy levels. Within this group, Ba ($Z=56$) dominates the Planck mean opacity at almost any temperature. However, this predominance gradually decreases with increasing temperature, in aid of Fr ($Z=87$).


Among the lanthanides (see upper panel of Figure~\ref{fig:lanthanides_actinides}) , for a temperature around $T=5000$ K, Sm ($Z=62$) has the largest expansion opacities at almost any wavelengths followed by Nd ($Z=60)$ and Pr ($Z=59$), while Dy ($Z=66$) opacity is also significant at low wavelengths. Sm expansion opacity remains the dominant one at any typical temperature of the KN ejecta 1 day post-merger ($1000$ K $\leq T \leq 10000$ K). In contrast, Lu ($Z=71$) has the smallest contribution to the expansion opacity for almost any temperature and wavelength. As far as actinides are concerned (see lower panel of Figure~\ref{fig:lanthanides_actinides}), U ($Z=92$) and Np ($Z=93$) expansion opacities are the highest ones nearly independent of the wavelength and temperature, followed by Pu ($Z=94$), the latter even becoming dominant at higher temperatures along with Am ($Z=95$). Overall, it is clear that lanthanide and actinide opacities are larger by several orders of magnitude with respect to all the other elements. This was obviously expected in light of the complex atomic structures of lanthanides and actinides, which are characterized by very large numbers of atomic levels giving rise to extremely high spectral densities. As an example, we can cite the f-shell element Nd II, whose atomic structure (as modeled with \HFR\ using 26 configurations in this work) is characterized by more than 13 millions of lines (with $\log gf \geq -5$), whereas the 40 configurations included in the CI model of 4d element Re I gives rise to about half less transitions, despite the much larger number of configurations considered. In addition, it is worth noticing that actinide opacities are generally larger than lanthanide ones for the homologous species (i.e. elements from the same column of the periodic table). In particular, the uranium expansion opacity is higher than the neodymium one for almost all temperatures and wavelengths within the ranges considered.

\begin{figure}[!htb]
    \centering
    \includegraphics[width=\hsize]{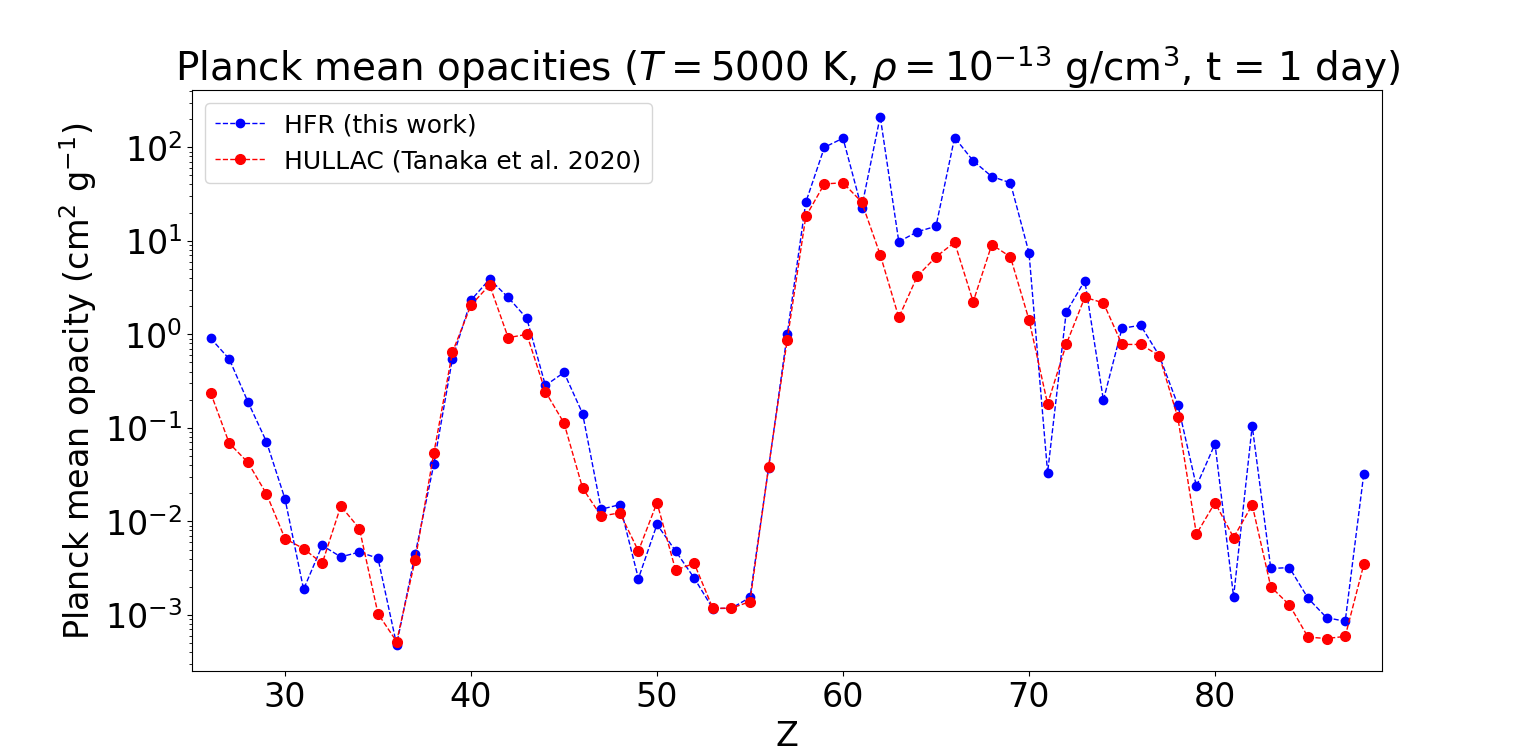}
    \caption{Comparison between Planck mean opacities obtained in this work and the ones from \citet{Tanaka2020} for all elements characterized by $Z =$ 26 -- 88, for $t=1$ day post-merger, $T=5000$K and $\rho=10^{-13}$ cm$^{-1}$.}
    \label{fig:Tanaka}
\end{figure}

A comparison of the Planck mean expansion opacities computed using the \HFR\ data obtained in this work with the ones obtained by \citet{Tanaka2020} using the HULLAC code \citep{Barshalom2001,Gu2008} is displayed in Figure \ref{fig:Tanaka}, for typical conditions expected within the KN ejecta one day after the NSM. The trend observed in this figure for the Planck opacities computed by both methods seems to reasonably agree to each other for non-lanthanide elements \citep[][did not compute actinide opacities]{Tanaka2020}, although our \HFR\ values tend to be larger in many cases. However, the Planck mean opacities of almost all lanthanide species (with a few exceptions) computed in this work are found to be significantly larger (by up to one order of magnitude) than the ones published by \citet{Tanaka2020}. This difference has already been commented and explained for Nd in one of our previous works on this topic \citep{Flors2023}, and is due to several aspects. Firstly, \HFR\ and HULLAC methods are based on rather different approaches. Relativistic corrections are introduced in \HFR\ through the pseudo-relativistic approach with $j$-independent radial wave functions, while in HULLAC the atomic relativistic states are obtained from the many electron Dirac Hamiltonian. However, the states from all configurations are optimized in the former, whereas the optimization of the potential is based on a few number of configurations in the latter. In addition, the configuration models used by \citet{Tanaka2020} are restricted to small numbers of configurations with respect to the CI models considered in our \HFR\ calculations, resulting in a smaller number of transitions considered and, thus, to a possible underestimation of the corresponding opacities. As an example, for \ion{Pr}{II}, only 6 configurations were considered by \citet{Tanaka2020} in their HULLAC calculation, while 26 configurations were included in our \HFR\ model based on an expansion opacity convergence study \citep{Flors2023}. Besides, as we recently highlighted in \citet{Carvajal2023c}, the partition functions used to compute the Sobolev optical depth (see Eq.~(\ref{equ:sobolev_optical_depth})) by \citet{Tanaka2020} were not computed using all the energy levels from the calculation but were approximated to the statistical weight of the ground level only, as shown in Eq. (7) in a previous work of this group \citep{Gaigalas2019} and mentioned in \citet{Flors2023}, or to a sum of the statistical weights up to a certain energy level (thus independent of the temperature) as detailed in \citet{Banerjee2024}. We showed in \citet{Carvajal2023c} that such approximation can have a significant impact on the computed expansion opacities, which could also explain some of the differences observed. \citet{Banerjee2024} confirmed this statement, showing a significant effect of such approximation on the expansion opacity of a lanthanide species (Eu), while lower-$Z$ element opacities seem less affected by the use of such approximated partition function. It is worth highlighting that the larger opacities obtained when using \HFR\ atomic data compared to the ones from \citet{Tanaka2020} using HULLAC, as well as the opacities determined with FAC data in \cite{Flors2023}, have already been discussed in the latter paper, the main reasons being found to be the higher density of levels near the ground state in the \HFR\ computations \citep{Flors2023}, as well as the extra configurations added in the \HFR\ multiconfigurations models and the different optimization approaches of the various methods, as already discussed above.

\section{Line-binned opacity}
\label{sec:LB}
In addition to expansion opacities, another frequency-dependent opacity formalism is the so-called line-binned opacity \citep{Fontes2020}: 
\begin{equation}
\kappa^{\rm bin}_{\nu} = \frac{1}{\Delta \nu}\frac{\pi e^2}{\rho m_e c}
\sum_{l} N_l
\, f_{l} \,,
\label{equ:line_binned_opacity}
\end{equation}
where $\rho$ is the ejecta density, $\Delta \nu$ are the widths of the frequency bins that contain lines $l$ with number densities of the lower level $N_l$ and oscillator strengths $f_l$. This expression is obtained by replacing the line profile with a flat distribution across the corresponding bin \citep{Fontes2020}. Unlike the expansion opacity (Eq.~(\ref{equ:expansion_opacity})), the line-binned opacity is independent of the evolution time, $t$, and it can therefore be used for on-the-fly table interpolation in a more straightforward manner.

\begin{figure}[!htb]
    \centering
    \includegraphics[width=\hsize]{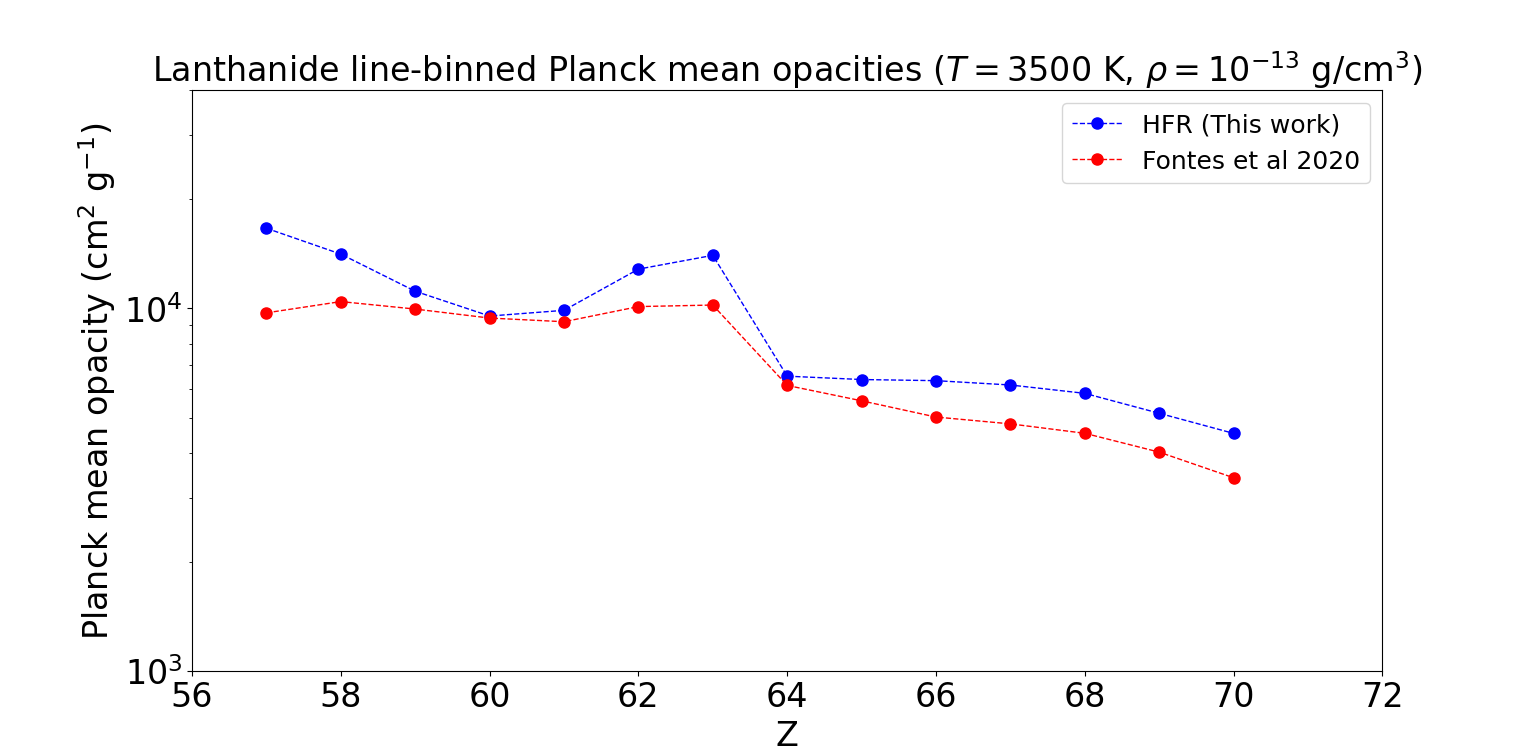}
    \caption{Lanthanide Planck line-binned opacities for $T=3500$ K and $\rho=10^{-13}$ cm$^{-1}$ in comparison with the one computed by \citet{Fontes2020}.}
    \label{fig:lanthanides_LB_Planck}
\end{figure}

\begin{figure}[!htb]
    \centering
    \includegraphics[width=\hsize]{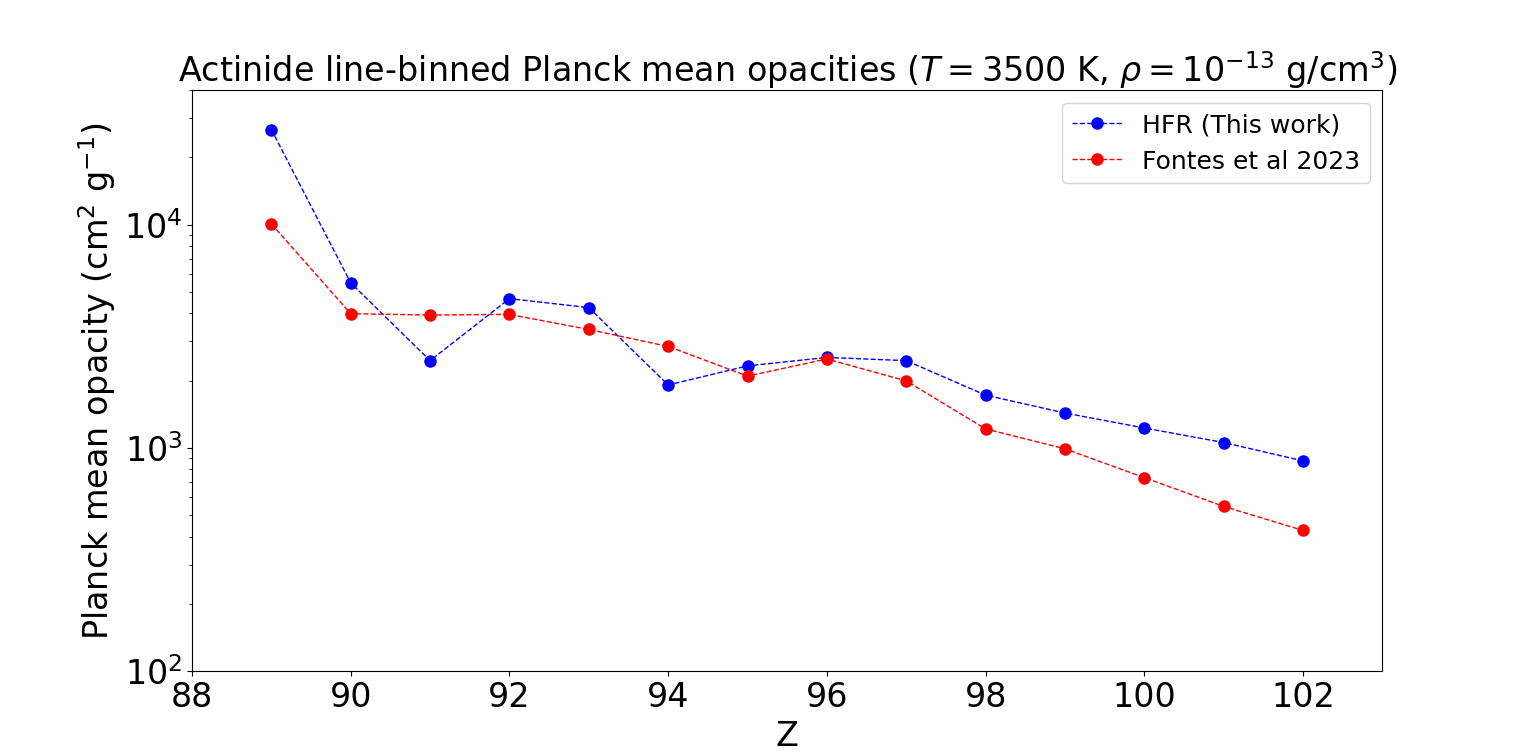}
    \caption{Actinide Planck line-binned opacities for $T=3500$ K and $\rho=10^{-13}$ cm$^{-1}$ in comparison with the one computed by \citet{Fontes2023}.}
    \label{fig:actinides_LB_Planck}
\end{figure}

We thus also computed the line-binned opacities in addition to the expansion ones for lanthanide and actinides species, in order to compare with other works, as well as their corresponding Planck mean opacities (as defined by Eq. (\ref{eq:planck})). In particular, we compared our results with the Planck mean line-binned opacities determined by \citet{Fontes2020} and \citet{Fontes2023} for lanthanides (Fig. \ref{fig:lanthanides_LB_Planck}) and actinides (Fig. \ref{fig:actinides_LB_Planck}), respectively. The Planck mean line-binned opacities that we computed for all lanthanides and actinides seem to agree in a reasonable extent to the ones from \citet{Fontes2020,Fontes2023}, even if our values are systematically slightly higher for lanthanides, with the same conclusion for actinides except for Pa ($Z=91$) and Pu ($Z=94$). The differences observed are thought to mainly arise from the configurations included in the atomic models.   

\section{HFR Atomic Database and Opacity Tables for Kilonovae}
\label{sec:database}

In this work, we created a database called "HFR Atomic Database and Opacity Tables for Kilonovae from Mons and Brussels Universities" which includes, for all the elements considered in this work (i.e. from $Z=20$ to $Z=103$), all the relevant atomic data to compute expansion and line-binned opacities in the context of kilonova emissions, as well as expansion and line-binned opacity tables for a grid of typical kilonova ejecta conditions. The atomic database contains all the energy levels and lines of each element considered in this work computed with the \HFR\ method. As for the expansion and line-binned opacity tables, they are provided for a grid of time post merger, density and temperature defined as follows: $t=$ 1, 2, 3, 4, 5, 6, 7 days ; $\rho = 10^{-17}, 10^{-16}, 10^{-15}, 10^{-14}, 10^{-13}$ g/cm$^{3}$ ; and $T=$ 1000, 2000, 3000, 4000, 5000, 6000, 7000, 8000, 9000, 10000 K. A table giving the Planck mean opacities of all these elements for the above-mentioned grid of conditions is also supplied. This database is available online on Zenodo at the address \url{https://zenodo.org/records/14017953} (DOI 10.5281/zenodo.14017952).

\section{KN ejecta opacity and light curves}
\label{sec:astro}

To illustrate the impact of the newly determined HFR opacities, we now consider a state-of-the-art NS merger simulation with the associated post-processed nucleosynthesis calculation, as described in \citet{Just2023}. More specifically, we consider the hydrodynamical simulations of a 1.375\Msun--1.375\Msun\ NS merger calculated with the SFHo EoS \citep{Just2023}. The total mass of 2.75\Msun\ corresponds to the estimated total mass of GW170817 event \citep{Abbott17}. This end-to-end model, referred to as ``sym-n1-a6'', includes the prompt dynamical ejecta component, the ejecta from the NS-torus system formed right after the merger, as well as the subsequently launched ejecta from the black-hole torus before its final evaporation. The masses of the three components amount to 0.006, 0.02 and 0.047 \Msun, respectively, and are sampled by 2380, 375 and 1180 mass elements (also referred to as ``trajectories''), respectively. 
The r-process calculations are performed using the BSkG3 mass model \citep{Grams23}, the associated TALYS reaction rates obtained with microscopic inputs for the nuclear level densities and photon strength functions \citep{Koning23}, $\beta$-decay rates from the RMF+RRPA model \citep{Marketin16}, the fission rates based on the HFB-14 model \citep{Goriely10a} and the fission fragment distributions obtained within the SPY model \citep{Lemaitre21}. 

\begin{figure}[!htb]
    \centering
    \includegraphics[width=\hsize]{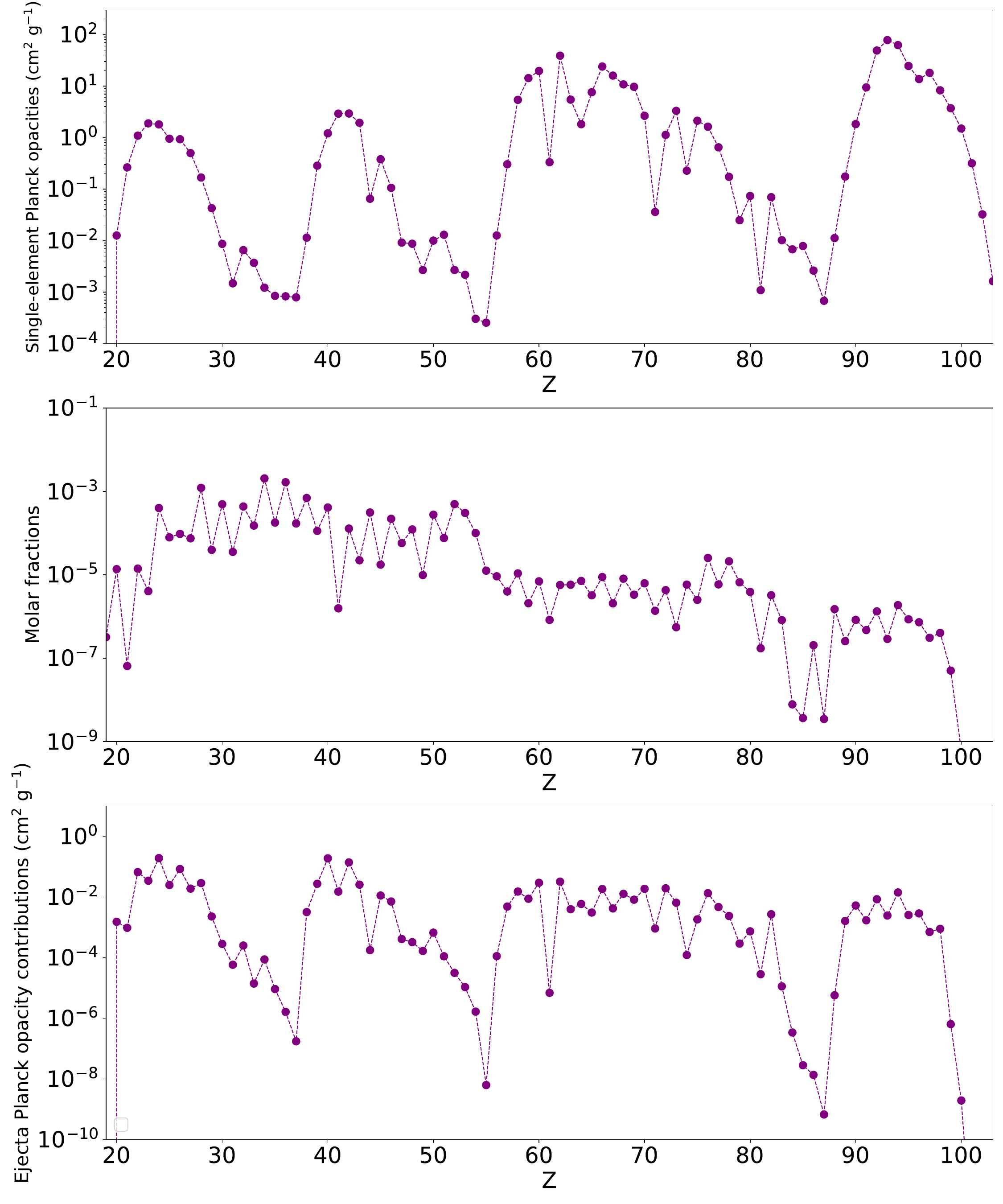}
    \caption{Single-element Planck mean opacities for each species at $t=3.5$~d, $T=6000$~K and $\rho=10^{-13}$~g/cm$^{3}$ (upper panel), elemental abundances ejected in the sym-n1-a6 NSM model at $t=3.5$~d \citep{Just2023} (middle panel), and contribution of each element to the Planck mean opacities of the KN ejecta based on the composition resulting in model sym-n1-a6 model 3.5 days post-merger (lower panel). The total Planck opacity of the ejecta (i.e. the HFR Planck mean opacities for this ejecta composition summed over all elements $Z$) is found to be equal to 1.13 cm$^{2}$g$^{-1}$, whereas an opacity of 1.43 cm$^{2}$g$^{-1}$ is obtained when using the formula from \citet{Just22a}}. 
    \label{fig:opave_symn1a6}
\end{figure}

The elemental molar fractions of the $7.3 \times 10^{-2}$ \Msun\ ejected in this model are shown in Fig.~\ref{fig:opave_symn1a6} (middle panel) as a function of the atomic number $Z$ for the entire ejecta 3.5 days after the NS merger. The contributions of all elements to the Planck mean expansion opacity of the ejecta at $t=3.5$~d, $T=6000$~K and $\rho=10^{-13}$~g/cm$^{3}$ are shown in Fig.~\ref{fig:opave_symn1a6} (lower panel). The partial opacities in this mixture of elements are determined by re-computing the expansion opacity (and, then, Planck mean expansion opacity) of each element by replacing the single-element transition lower level number densities $n_l$ independent of $Z$ (as defined in Eq. \ref{equ:sobolev_optical_depth}) by $Z-$dependent lower level number densities $n_{l,Z}$, which are obtained by multiplying the single-element number densities $n_l$ by the corresponding elemental abundance of each element, $y_Z$, i.e.
\begin{equation}
n_{l,Z} = y_Z n_l\,.
\end{equation} 
For this NS merger model, a major contribution to the total abundance-averaged opacity (for a composition averaged over the entire ejecta) comes from the 3d-shell $Z\simeq 24$ and 4d-shell $Z\simeq 40$ elements, in addition to the lanthanides and actinides. The lower opacity of the $Z\simeq 24$ and $Z\simeq 40$ elements is compensated by their relatively large abundances. As a consequence, lanthanides do not fully dominate the opacity, at least on average. The ejecta total HFR opacity (equal to 1.13 cm$^{2}$g$^{-1}$) is found to be about 25 \% lower than the values that would result from the heuristic prescription for the opacities deduced in \citet{Just22a}. In that approach, the opacity is parametrically expressed as a function of the lanthanide plus actinide molar fractions (which is equal to $8.4 \times 10^{-5}$ in this model) and the gas temperature (see Eqs. 13--15 from \citet{Just22a}). In addition, it is worth mentioning that estimating the contribution of each element to the ejecta Planck expansion opacity by averaging the single-element Planck opacities (as shown in Fig.~\ref{fig:opave_symn1a6}; upper panel) with the elemental abundances leads to a rather poor approximation, since a corresponding value of an ejecta total Planck opacity of 0.11 cm$^{2}$g$^{-1}$ would be found in this case, i.e. one order of magnitude lower than the actual value. While most works report single-element expansion opacities (i.e. assuming a composition made of 100 \% of a given element for each species), the latter can therefore not be used to properly estimate the total opacity for a given composition a posteriori, although they are obviously important to compare results between various atomic physics approaches. This is not the case for line-binned opacities, which are directly proportional to the transition lower level number densities (see Eq. \ref{equ:line_binned_opacity}), hence to the elemental abundances. 

\begin{figure}[!htb]
    \centering
    \includegraphics[width=\hsize]{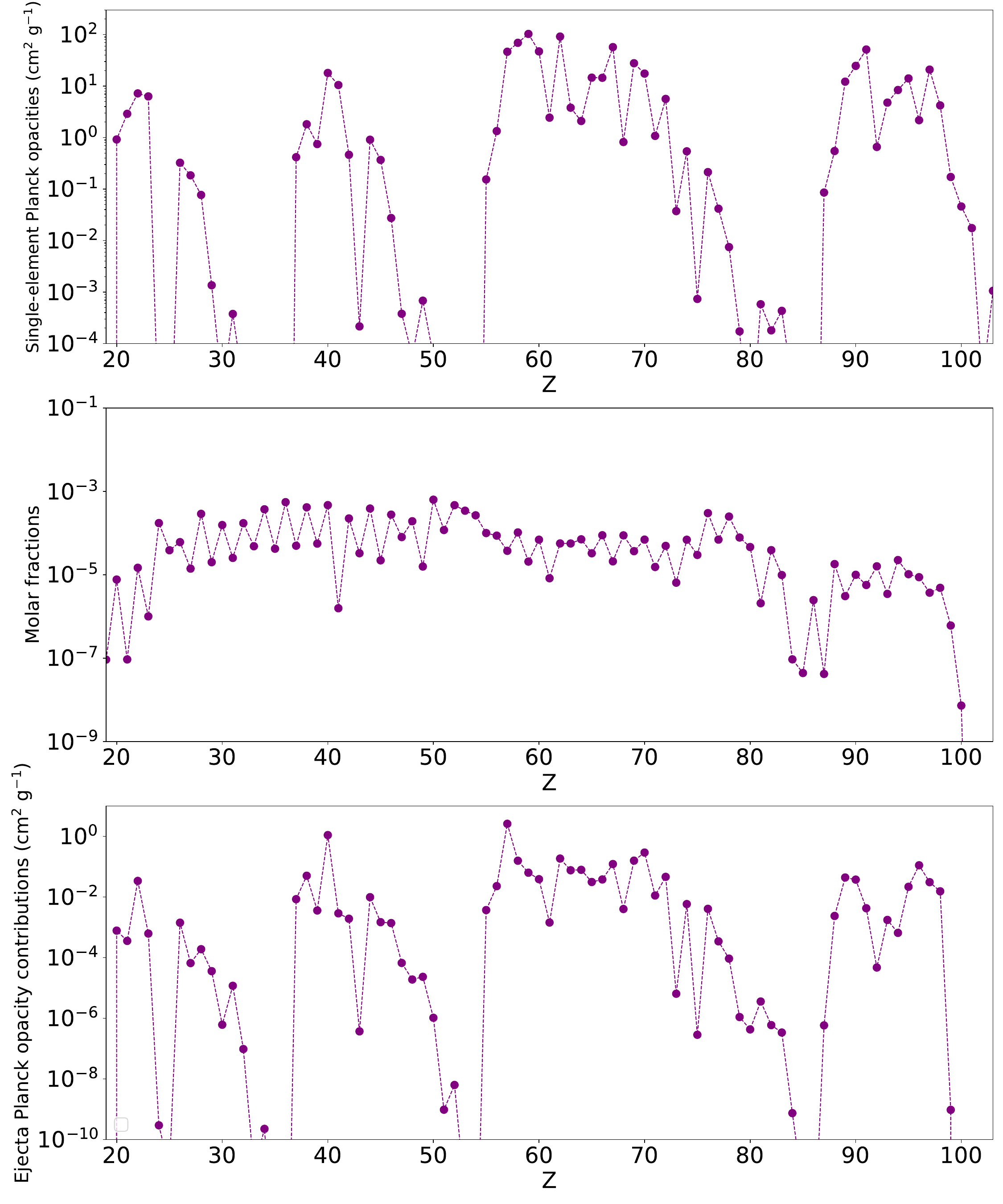}
    \caption{Same as Fig.~\ref{fig:opave_symn1a6} but only for the dynamical ejecta at $t=3.5$~d, $T=2500$~K and $\rho=10^{-15}$~g/cm$^{3}$. In this case, the total HFR opacity is equal to 5.42 cm$^{2}$g$^{-1}$, while the parametric formula from \citet{Just22a} gives an opacity of 2.87 cm$^{2}$g$^{-1}$.
    }
    \label{fig:opave_symn1a6_dyn}
\end{figure}

Similarly, Fig.~\ref{fig:opave_symn1a6_dyn} gives the single-element Planck mean expansion opacities, the molar fractions and the Planck opacity contributions of all elements just for the rapidly expanding dynamical component of the ejecta in the sym-n1-a6 model. In this case, the ejecta is relatively more abundant in lanthanides and actinides (their molar fraction is equal to $8.7 \times 10^{-4}$, i.e. one order of magnitude higher with respect to the entire ejecta case), but also $Z\simeq 40$ isotopes. The contribution of these three groups of elements are seen to dominate the total abundance-averaged opacity (see Fig.~\ref{fig:opave_symn1a6_dyn}, lower panel). In this specific case, at 3.5 days post merger, the phenomenological parametrisation of \citet{Just22a} underestimates the dynamical ejecta opacity by a factor of two, since the HFR total opacity is found to be of 5.42 cm$^{2}$g$^{-1}$, while an opacity of 2.87 cm$^{2}$g$^{-1}$ is obtained when using the formula from \cite{Just22a}.

\begin{figure}[!htb]
    \centering
    \includegraphics[scale=0.25]{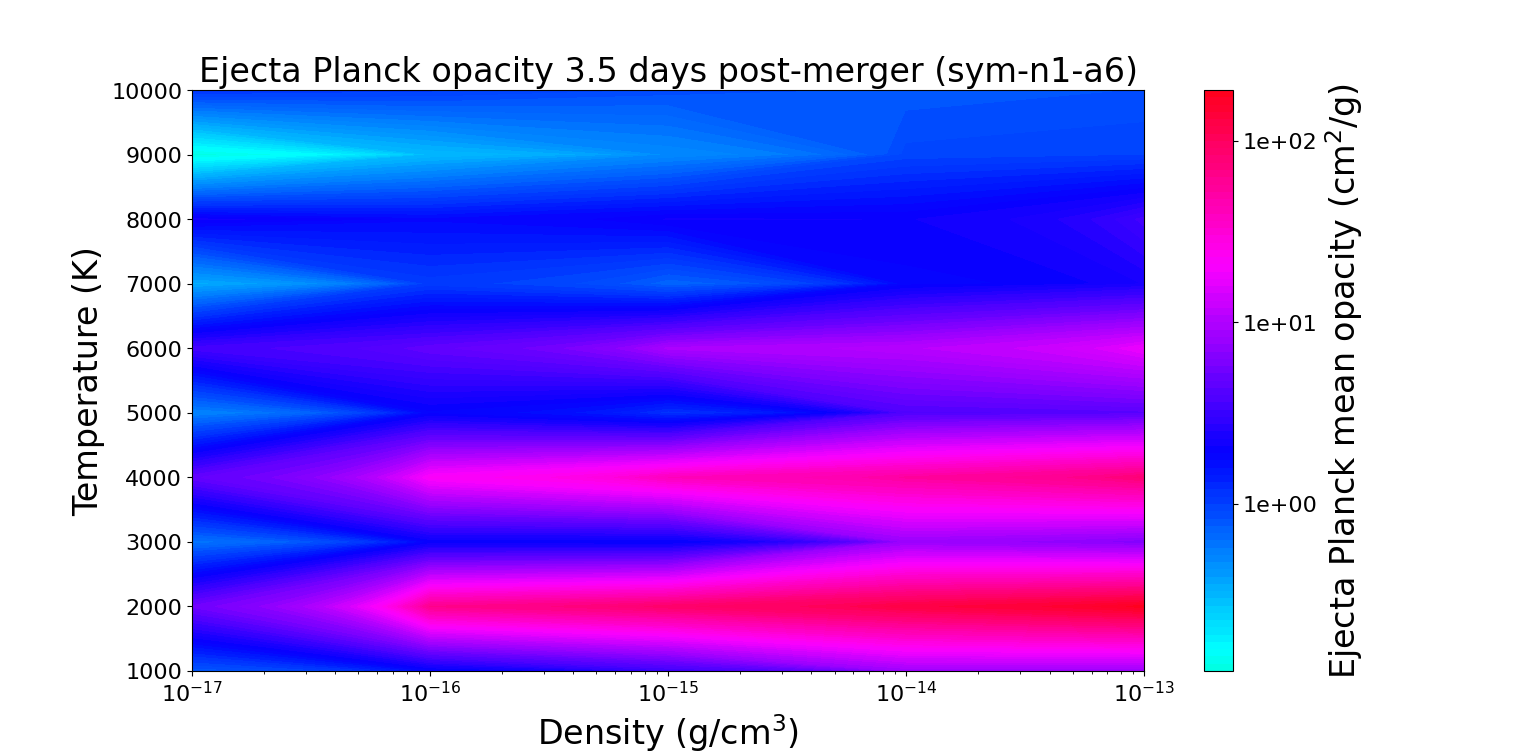}
    \caption{Temperature and density dependence of the Planck mean opacity of the ejecta at a time $t=3.5$ days post merger assuming the composition of model sym-n1-a6 at that time.
    }
    \label{fig:contour-plot}
\end{figure}

The dependence of the Planck mean opacity of the ejecta with respect to the temperature and density at a time $t=3.5$ days post-merger is illustrated in Fig. \ref{fig:contour-plot}, still considering the overall ejecta composition deduced from NS merger model sym-n1-a6 at $t=3.5$~d. It is clear from this figure that the ejecta opacity tends to increase with higher values of density and temperature, although non-monotonic behavior is also seen, particularly in the temperature dependence, which shows a kind of band structure. As low ionization stages dominate at low temperature, and since energy levels lie closer in lowly-charged ions, the ionization structure can explain the opacity raise for decreasing temperatures and the resulting band structure observed.

\begin{figure*}[htb!]
    \centering
    \begin{subfigure}[t]{0.45\textwidth}
        \centering
        \includegraphics[height=3in]{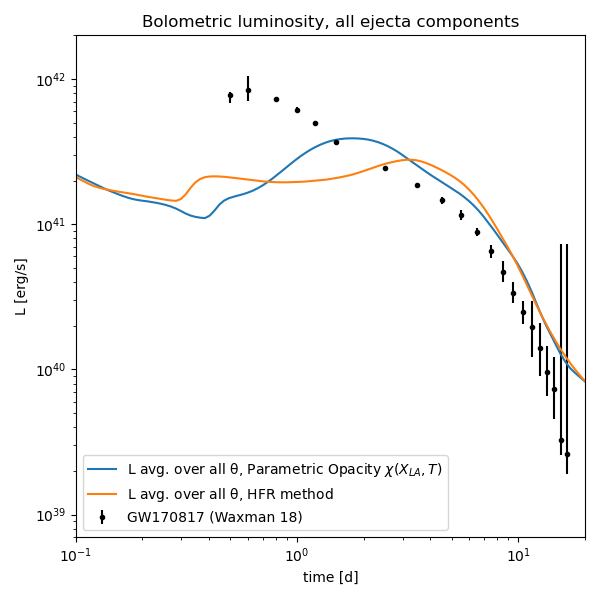}
        \caption{}
    \end{subfigure}
    \begin{subfigure}[t]{0.45\textwidth}
        \centering
        \includegraphics[height=3in]{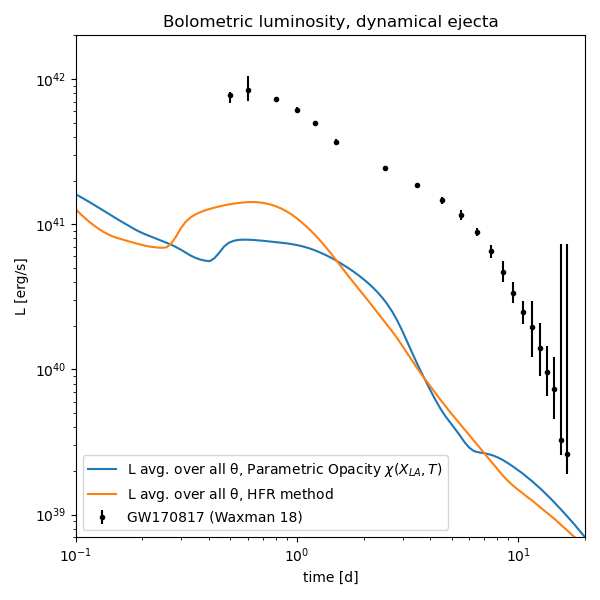}
        \caption{}
    \end{subfigure}
    \caption{Comparison of light curves obtained with the kilonova scheme developed in \cite{Just22a} for the NS merger model sym-n1-a6, using a parametric opacity function defined by \cite{Just22b} or the opacity data calculated using the HFR method. The left panel shows the comparison for all the ejecta components including the prompt dynamical component, the ejecta from the NS-torus system, and the disk wind stemming from the black-hole torus remnant; while the right panel shows the comparison for the dynamical ejecta component alone. Note that the computed light curves adopt the simplified assumption of spatially averaged composition. The observed bolometric luminosity of AT2017gfo \citep{Waxman18} is shown by solid circles for reference.}\label{fig:Alcar_results}
\end{figure*}

\begin{figure}[!htb]
    \centering
    \includegraphics[width=\hsize]{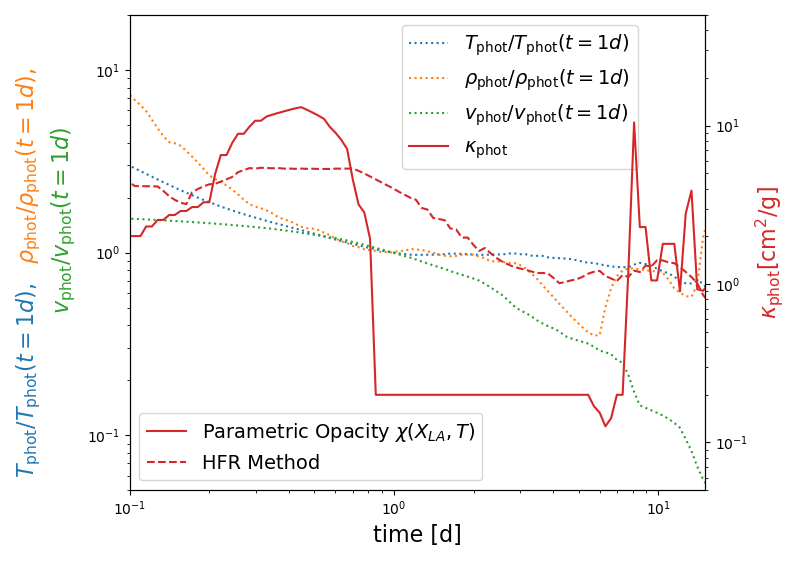}
    \caption{Time evolution of the photospheric properties and the opacity calculated using the HFR method presented in this paper for the NS merger model sym-n1-a6 and assuming a uniform composition corresponding to the full ejecta at $t=3.5$~d. Temperature (blue, dotted line), density (orange, dotted line), and velocity (green, dotted line) are scaled by the left y-axis and normalized by their values at 1 day. The values at 1 day for these quantities are $T_{ph}(t=1\rm{d})=2750$ K, $\rho_{ph}(t=1\rm{d})=7.4 \times 10^{-16}$ g cm$^{-3}$, and $v_{ph}(t=1\rm{d})=1.2 \times 10^{10}$ cm s$^{-1}$, respectively. The Planck-averaged opacity for the HFR data (red, dashed line) is plotted on the right y-axis. The solid red curve shows the opacity calculated using the parametric opacity function defined by \cite{Just22b}, for comparison.}
    \label{fig:ph_prop_symn1a6}
\end{figure}

The calculated yields and radioactive heating rates are now used as input for a two-dimensional (axisymmetric) radiation transport simulation adopting approximate M1 closure to estimate the KN light curve, as detailed in \citet{Just22a}. The results are shown in Fig.~\ref{fig:Alcar_results}. In panel (a), a comparison of the bolometric light curves is shown for the parametric opacity function defined in \citet{Just22a} as well as for the case using the opacities calculated by the HFR method described in this paper. 
For the calculation using the HFR opacities, we adopt the average composition as shown in Fig.~\ref{fig:opave_symn1a6} and assume that the ejecta composition is uniformly distributed in the 2D plane. All of the ejecta components are included in this model. The panel (b) shows the same comparison but assuming that only dynamical ejecta are present, by setting the densities of all other ejecta components to zero and assuming the averaged composition as shown in Fig.~\ref{fig:opave_symn1a6_dyn} to be homogeneously distributed in the ejecta. Note that the ejecta composition is not averaged for the simulation with the parametric opacity function. The observed bolometric luminosity of AT2017gfo \citep{Waxman18} is also displayed by solid circles for reference. Using the HFR opacities makes a noticeable impact on the light curve compared to the parametric ones. The HFR opacity yields a slightly lower peak bolometric luminosity when all the ejecta components are included, and the time of the peak is shifted toward $\sim$7-8 days, as compared to $\sim$1-2 days for the simple parametric opacity approximation of \citet{Just22b}. We illustrate in Fig. \ref{fig:ph_prop_symn1a6} the time evolution of the averaged photospheric properties and the opacity for the ALCAR-kilonvoa model using the \HFR \ data for the entire ejecta. The photospheric temperature, density, and velocity normalized by their values at 1 day are shown. The Planck mean opacity for both the \HFR \ data and for the parametric opacity function are compared. As seen in this figure, the opacity is initially higher for the parametric function as compared to the \HFR \ opacity until about 1 day, when it drops significantly. The \HFR \ opacity drops more steadily and reaches the minimum value at around $\sim$7-8 days. This is consistent with the behavior of the model light curves in Fig.~\ref{fig:Alcar_results}a. When only the fast dynamical ejecta component is included, the contribution to the opacity is dominated by lanthanides and actinides. In this case, the light curve is comparable for both the opacity calculations, but the peak luminosity is higher for \HFR \ data as compared to the parametric opacity approximation. These results illustrate that opacities based on atomic data are sorely needed in order to reliably predict KN light curves.

\section{Conclusions}
This work consists in a new large-scale calculation of atomic parameters and opacities for all elements from Ca ($Z=20$) to Lr ($Z=103$) in their first four charge stages, with a special focus on lanthanides and actinides. This study is carried out in the context of kilonovae, the electromagnetic counterpart of gravitational wave emission following the coalescence of two NSs, a relevant site of heavy-element production by the r-process. The purpose of this work is to is to build a database providing more accurate values for the kilonova ejecta opacity based on detailed atomic calculations. These opacities are of great impoortance for modeling kilonova light curves and spectra. 

Atomic computations have been performed using the pseudo-relativistic Hartree-Fock (\HFR) method, which has already been used in our group, in the same context, for \ion{Ce}{II--IV} \citep{Carvajal2021}, for \ion{Nd}{II--III} and \ion{U}{II--III} \citep{Flors2023, Deprince2023}, for \ion{Er}{III} \citep{Deprince2024}, for several lighter trans-iron elements \citep{BenNasr2023,BenNasr2024}, as well as for many moderately-charged lanthanide ions \citep{Carvajal2022a, Carvajal2022b, Carvajal2023a, Carvajal2023b, Maison2022}. The strategy adopted in this work is to build the atomic multiconfiguration models of all elements in the charge stages\ion{}{I--IV} (which are the species expected to be present in the KN ejecta). Our approach is justified based on the methodology developed in our previous studies for selected lanthanides and actinides \citep{Flors2023, Deprince2023, Deprince2024} as well as the one used for several lighter trans-irons elements \citep{BenNasr2023,BenNasr2024}, in which we also compare with available experimental data as well as with results obtained using another computational method \citep[namely, FAC, ][]{Gu2008}.

The expansion opacities computed for all elements from Ca to Lr using our \HFR\ atomic data are illustrated in this article for typical conditions expected in the KN ejecta one day after merger (namely, at a fiducial temperature $T=5000$ K and a density $\rho = 10^{-13}$ cm$^{-3}$). The corresponding Planck mean opacities are estimated for a grid of temperatures ranging from $T=1000$ K to $T=10000$ K. For such physical conditions within the ejecta, we confirm that lanthanide and actinide opacities are significantly higher than the other trans-iron element ones, which can be explained by the complex atomic structures of these ions characterized by an unfilled 4f and 5f subshell, respectively. Among lanthanides, we note that Sm, Nd and Pr have the largest opacities, while U, Np and Pu dominate the actinide expansion opacities. A comparison with the Planck mean expansion opacities determined using HULLAC atomic data for elements with $20 \leq Z \leq 88$ by \citet{Tanaka2020} is also discussed in details. From the latter, we conclude that the trends from both calculations agree to a good extent, although \HFR\ Planck mean expansion opacities tend to be larger for several ions, especially for lanthanides by up to one order of magnitude. This disagreement can be explained by the differences between both methods, but especially by the differences in the multiconfiguration models used in both computations, \citet{Tanaka2020} using more restricted atomic models as well as approximated atomic partition functions \citep[see][for more details]{Carvajal2023c,Banerjee2024}. In addition, we find that the Planck mean line-binned opacities determined in this work for both lanthanides and actinides are in reasonable agreement with those obtained by \citet{Fontes2020,Fontes2023} (though most of our \HFR\ line-binned opacities are found to be systematically slightly higher).

It is also worth highlighting that we published a database called "HFR Atomic Database and Opacity Tables for Kilonovae from Mons and Brussels Universities" including all the relevant data to compute expansion and line-binned opacities in the context of kilonova emissions, as well as expansion and line-binned opacity tables for a grid of typical kilonova ejecta conditions.

Finally, we applied our \HFR\ Planck mean expansion opacities to calculate KN light curves using the ALCAR-kilonova code developed in \citep{Just22a} for a 1.375 \Msun-1.375 \Msun\ NS merger model \citep{Just2023}. 
We found that the phenomenological ejecta opacity used by \citet{Just22a} can be significantly different from the one determined using the present \HFR\ atomic data.
In addition to lanthanides and actinides, a non-negligible contribution to the opacity comes from the abundant elements around $Z \simeq 24$ and 40. Lanthanides thus seem to not fully dominate the ejecta opacity, at least on average.
The \HFR\ opacities yield a lower luminosity for the first week and shift the peak luminosity to a later day (around 7-8 days) when compared to the phenomenological prescription of \citet{Just22a} (which peaks around $\sim$2 days). 
When considering the fast dynamical ejecta only, which are dominated by the abundant f-shell lanthanides and actinides, the phenomenological formula from \citet{Just22a} gives a rather similar light curve for the first 10 days, but the HFR opacities yield a brighter peak. 
The added value of the phenomenological prescription is that the spatial and temporal dependence of the composition can be taken into account in the calculation of the opacity, something that cannot be easily done with microscopic composition-dependent expansion opacities at the moment.
Obviously, opacities based on detailed atomic data are sorely needed to reliably predict light curves. 
In the present simulations, a uniform composition of the ejecta in the 2D plane has been assumed due to the complex dependence of the expansion opacity on the number densities. An important future improvement in the estimate of KN light curves would consist in calculating the opacities along the ejecta trajectories according to the specific composition of each cell of the 2D grid. Since this calculation represents a significant computer effort, we leave it to future work.

In view of the present results, it would also be worth further improving the \HFR\ atomic data for the species that are found to contribute the most to the kilonova ejecta opacity, among which feature several lanthanides and actinides (in particular Sm, Nd, Dy, Er, U, Np and Pu) as well as some lighter trans-iron elements around $Z \simeq 24$ and $Z \simeq 40$ (such as Cr, Fe, Zr, Mo). For these elements, more elaborate atomic models could be tested, \textit{e.g.} enlarging the configuration lists, or including the so-far neglected core-polarization correction. Fits to experimental energy levels, when available, could also be attempted in a second step to assess the accuracy of our data. An extended theory-observation comparison of atomic energy levels and line intensities (see e.g. \citet{Ding2024} for \ion{Nd}{III}) might indeed allow some semi-empirical adjustment of our computed \HFR\ atomic parameters and hence improve the resulting opacities.

\begin{acknowledgements}
The present work is supported by the FWO and F.R.S.-FNRS under the Excellence of Science (EOS) programme (numbers O.0004.22 and O022818F). HCG is a holder of a FRIA fellowship. SG and PP
are Research Associate of the Belgian Fund for Scientific
Research F.R.S.-FNRS. PQ is F.R.S.-FNRS Research Director. OJ acknowledges support by the European Research Council
(ERC) under the European Union’s Horizon 2020 research and innovation programme (ERC Advanced Grant
KILONOVA No. 885281), by the Deutsche Forschungsgemeinschaft (DFG, German Research Foundation) through Project - ID 279384907 – SFB 1245 (subprojects B06, B07), and by the State of Hesse within the Cluster Project ELEMENTS. Computational resources have been provided by the Consortium des Equipements de Calcul Intensif (CECI), funded by the F.R.S.-FNRS under Grant No. 2.5020.11 and by the Walloon Region of Belgium. We also want to thank our collaborators from GSI, Universidade de Lisboa and NOVA University of Lisbon for fruitful discussions, in particular A. Flörs, R. Silva, J. Marques, G. Martínez-Pinedo and J. Sampaio.
\end{acknowledgements}

\bibliographystyle{aa}
\bibliography{paper}

\newpage  

\section*{Appendix}


    \onecolumn
\begin{center}

    \topcaption{List of the configurations used to model each atomic system}
\begin{supertabular}{p{1cm} p{15cm}}

    \label{table:config}
 
Ca I & 4s$^{2}$, 4s5s, 4s6s, 4s3d, 4s4d, 4s5d, 4s6d, 3d$^{2}$, 4p$^{2}$, 4s4p, 4s5p, 4s6p, 4s4f, 4s5f, 4s6f\\ Ca II    & 4s, 5s, 6s, 4d, 5d, 6d, 4p, 5p, 6p, 4f, 5f, 6f\\ Ca III   & 3s$^{2}$3p$^{6}$, 3s$^{2}$3p$^{5}$4p, 3s$^{2}$3p$^{5}$5p, 3s$^{2}$3p$^{5}$6p, 3s$^{2}$3p$^{5}$4f, 3s$^{2}$3p$^{5}$5f, 3s$^{2}$3p$^{5}$6f, 3s$^{2}$3p$^{5}$4s, 3s$^{2}$3p$^{5}$5s, 3s$^{2}$3p$^{5}$6s, 3s$^{2}$3p$^{5}$3d, 3s$^{2}$3p$^{5}$4d, 3s$^{2}$3p$^{5}$5d, 3s$^{2}$3p$^{5}$6d\\ Ca IV    & 3s$^{2}$3p$^{5}$, 3s$^{2}$3p$^{4}$4p, 3s$^{2}$3p$^{4}$5p, 3s$^{2}$3p$^{4}$6p, 3s$^{2}$3p$^{4}$4f, 3s$^{2}$3p$^{4}$5f, 3s$^{2}$3p$^{4}$6f, 3s3p$^{5}$3d, 3s$^{2}$3p$^{4}$4s, 3s$^{2}$3p$^{4}$5s, 3s$^{2}$3p$^{4}$6s, 3s$^{2}$3p$^{4}$3d, 3s$^{2}$3p$^{4}$4d, 3s$^{2}$3p$^{4}$5d, 3s$^{2}$3p$^{4}$6d, 3s3p$^{6}$\\ Sc I     & 3d4s$^{2}$, 3d4s5s, 3d4s6s, 3d4s4d, 3d4s5d, 3d4s6d, 3d$^{3}$, 3d$^{2}$4s, 3d$^{2}$5s, 3d$^{2}$6s, 3d$^{2}$4d, 3d$^{2}$5d, 3d$^{2}$6d, 3d4p$^{2}$, 3d4p4f, 3d4d$^{2}$, 3d4f$^{2}$, 4s$^{2}$4d, 3d$^{2}$4p, 3d$^{2}$5p, 3d$^{2}$6p, 3d$^{2}$4f, 3d$^{2}$5f, 3d$^{2}$6f, 3d4s4p, 3d4s5p, 3d4s6p, 3d4s4f, 3d4s5f, 3d4s6f, 3d4p5s, 3d4p6s, 3d4p4d, 3d4p5d, 3d4p6d, 4s$^{2}$4p, 4s$^{2}$4f\\ Sc II    & 3d4s, 3d5s, 3d6s, 3d4d, 3d5d, 3d6d, 3d$^{2}$, 4s$^{2}$, 4s5s, 4s6s, 4s4d, 4s5d, 4s6d, 4p$^{2}$, 4p4f, 4d$^{2}$, 4f$^{2}$, 3d4p, 3d5p, 3d6p, 3d4f, 3d5f, 3d6f, 4s4p, 4s5p, 4s6p, 4s4f, 4s5f, 4s6f, 4p5s, 4p6s, 4p4d, 4p5d, 4p6d\\ Sc III   & 3d, 4d, 5d, 6d, 4s, 5s, 6s, 4p, 5p, 6p, 4f, 5f, 6f\\ Sc IV    & 3s$^{2}$3p$^{6}$, 3s$^{2}$3p$^{5}$4p, 3s$^{2}$3p$^{5}$5p, 3s$^{2}$3p$^{5}$6p, 3s$^{2}$3p$^{5}$4f, 3s$^{2}$3p$^{5}$5f, 3s$^{2}$3p$^{5}$6f, 3s$^{2}$3p$^{5}$3d, 3s$^{2}$3p$^{5}$4d, 3s$^{2}$3p$^{5}$5d, 3s$^{2}$3p$^{5}$6d, 3s$^{2}$3p$^{5}$4s, 3s$^{2}$3p$^{5}$5s, 3s$^{2}$3p$^{5}$6s\\
\end{supertabular}
\end{center}   

%
%

\end{document}